# Dual-pulse micronozzle acceleration of sub-GeV-class protons

D. Pan and M. Murakami[*]

*Institute of Laser Engineering, Osaka University, Suita, Osaka 565-0871, Japan*

We propose a dual-pulse micronozzle acceleration scheme that enables *phase-locked acceleration* of laser-driven protons, mitigating the conventional trade-off between maximum energy and laser-to-proton conversion efficiency. By introducing a delay-tuned synchronization window, a compact proton front generated by a shaping prepulse is injected into, and remains copropagating with, a quasistatic axial electric field driven by a delayed main pulse in a micronozzle cavity. This phase locking maintains the relative phase between the proton bunch and the accelerating field over an extended interaction length and duration, thereby suppressing thermal debunching and prolonging the effective acceleration stage. At main-pulse intensities of the order of $10^{21}\ \mathrm{W/cm^2}$, sub-GeV-class proton cutoffs are achieved with a total laser-to-proton conversion efficiency of ∼20%. Notably, the efficiency of the application-relevant high-energy component exceeds ∼13% for protons with energies $E_p > 100$ MeV, indicating preferential energy loading into a compact, directed proton population rather than quasithermal sheath expansion. Comparative simulations with an unconfined dual-pulse hydrogen-rod target demonstrate that this performance gain arises from the combined effects of temporal synchronization and geometric confinement, which sustain a long-lived axial accelerating channel advected downstream with the proton front. An analytical model for the synchronization condition is developed and validated against the simulation data. Three-dimensional particle-in-cell simulations support that the phase-locking mechanism and the associated spectral hardening are preserved in 3D slit-nozzle geometries, with the confined target yielding cutoff energies ∼60% higher than an unconfined hydrogen rod under identical laser conditions. These results identify phase-locked acceleration in a confined laser-driven structure as a practical design principle for compact, high-yield sub-GeV-class proton drivers, with direct relevance to secondary-particle sources such as pion and muon production, accelerator-driven systems, and laser-based neutron sources.

## I. INTRODUCTION

Achieving sub-GeV to GeV-class proton energies with high laser-to-particle conversion efficiency remains a central challenge in laser-driven ion acceleration [1,2]. Established mechanisms include target-normal sheath acceleration (TNSA) [3–5], radiation-pressure acceleration [6,7], collisionless shock acceleration [8], magnetic vortex acceleration [9], the breakout-afterburner [10,11], and Coulomb explosion [12,13]. These mechanisms can generate energetic ions but generally suffer from rapid dephasing between the accelerating field and the ion population, as well as transverse and longitudinal debunching during expansion [2,14]. As a result, increases in peak proton energy are typically accompanied by a sharp reduction in conversion efficiency, which poses a fundamental limitation for compact laser-based proton drivers intended for secondary-particle generation and accelerator applications [15,16].

Various strategies have been proposed to mitigate these limitations, such as tailoring target geometry, shaping laser temporal profiles, or employing multipulse configurations [17–22]. However, in most cases, the accelerating field either decays rapidly or propagates independently of the proton bunch, leading to an intrinsically dephasing-limited process. From this perspective, the central challenge is not injection itself, but the sustained preservation of phase coherence between the proton bunch and a comoving accelerating structure. Indeed, the importance of injecting ions into the correct phase of the accelerating field has been demonstrated in radiation-pressure and collisionless-shock contexts [23–27], underscoring the general role of phase locking in efficient ion acceleration.

Recently, Murakami *et al.* [28] introduced the micronozzle acceleration (MNA) concept, in which a solid-density aluminum nozzle provides transverse confinement and guides the laser-driven electron current to form a quasistationary accelerating channel. That study employed a single laser pulse and demonstrated that geometric confinement alone enhances proton energies relative to an unconfined target. However, it did not explore temporal control of the proton injection, nor did it address the interplay between pulse timing and the confined accelerating field.

[*]Contact author: murakami.masakatsu.ile@osaka-u.ac.jp

The present work extends the MNA concept in three directions. First, we introduce a dual-pulse (DP) scheme in which a tightly focused prepulse extracts a compact proton bunch from the hydrogen rod, and a delayed, larger-spot main pulse drives the quasistatic axial field inside the nozzle cavity (Fig. 1)]. The interpulse delay $\Delta t$ controls the spatial overlap between the proton front and the peak accelerating field, enabling *phase-locked* acceleration over an extended interaction length. Second, we develop an analytical model that predicts the optimal delay and the width of the synchronization window directly from the target geometry. Third, we present three-dimensional particle-in-cell (PIC) simulations that confirm the robustness of the phase-locking mechanism beyond the 2D approximation. The central new element of this work is the realization of phase-locked proton acceleration through delay-controlled synchronization in a confined micronozzle geometry.

We note that the solid-density aluminum (Al) nozzle employed here ($n_{\mathrm{Al}} = 6 \times 10^{22}\ \mathrm{cm}^{-3}$) is fundamentally different from the low-density gas nozzles and cluster jets studied in the early 2000s [29,30]. In those experiments, the Coulomb explosion of individual clusters at densities many orders of magnitude below solid provides the acceleration mechanism, whereas in the present scheme, the nozzle walls act as a rigid geometric boundary that confines the laser-driven electron current and sustains a collective axial field over hundreds of femtoseconds.

Finally, sub-GeV-class proton drivers with high energy efficiency are of direct interest for secondary-particle applications such as compact neutron and muon sources [31–33].

## II. RESULTS

The results below are obtained from fully relativistic PIC simulations performed with the EPOCH code [34]. We first develop an analytical model for the synchronization condition (Sec. II B), then present the 2D simulation results that verify phase-locked acceleration (Secs. II C and II D), demonstrate numerical convergence (Sec. II E), and finally validate the mechanism in three dimensions (Sec. II F).

### A. Simulation setup

To isolate the role of temporal control, we introduce only two minimal changes to the original MNA geometry [28] [Fig. 1]: (i) We replace the cylindrical H-rod with an elliptical rod of semiaxes $r_x = 2.8$ and $r_y = 1.4\ \mu\mathrm{m}$ in direct contact with the nozzle. (ii) We drive the target with a dual-pulse system and scan the intensities and durations of the prepulse and main pulse.

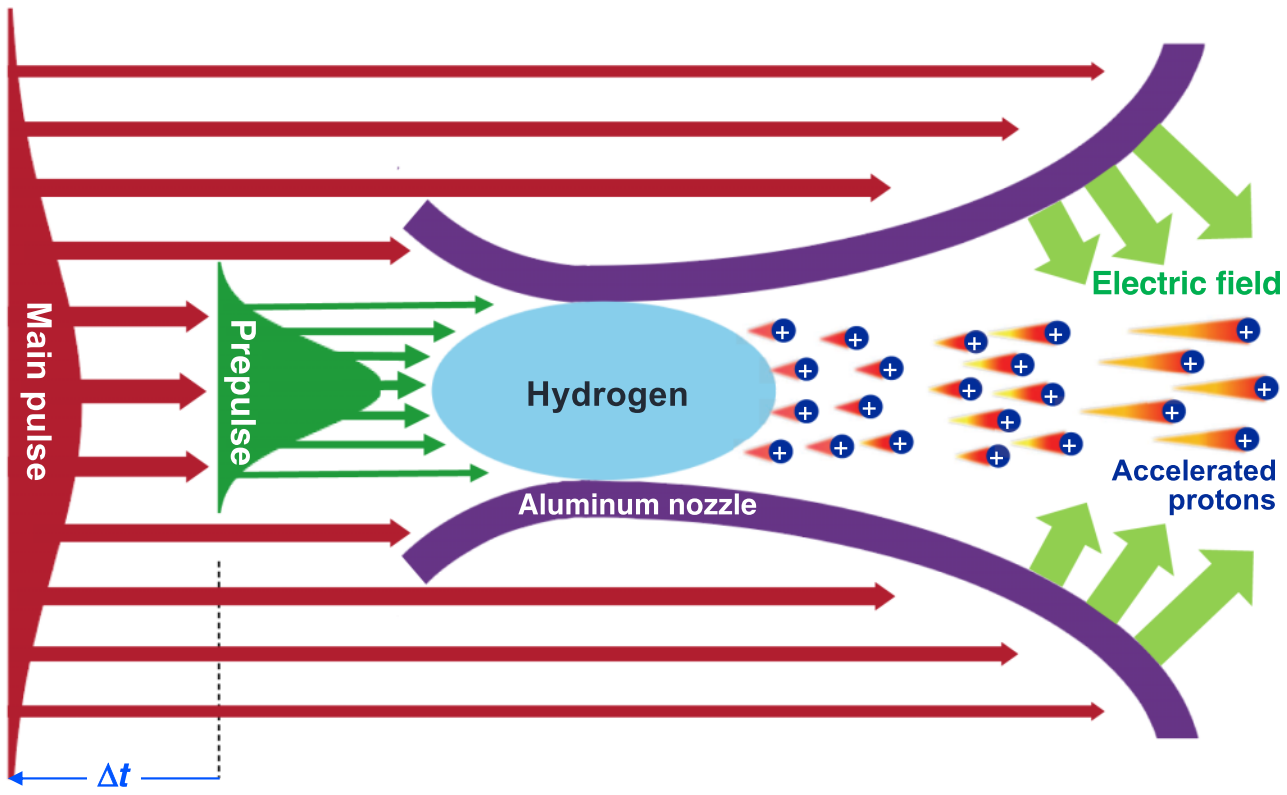

FIG. 1. Schematic of dual-pulse micronozzle acceleration (DP-MNA). A focused prepulse extracts a compact proton bunch from the hydrogen rod (H-rod); after a delay $\Delta t$, a larger-spot main pulse drives a quasistatic axial field in the Al nozzle, enabling phase-locked sub-GeV-class acceleration.

The spot sizes are fixed at $w_{\mathrm{main}} = 10\ \mu\mathrm{m}$ and $w_{\mathrm{pre}} = 2.8\ \mu\mathrm{m}$, and the delay $\Delta t \equiv t_{\mathrm{main}} - t_{\mathrm{pre}}$ is varied from $-80$ to $+80$ fs ($\Delta t > 0$: prepulse arrives first). The two pulses play complementary roles: a tightly focused prepulse shapes the proton source at the H-rod, while the delayed, larger-spot main pulse deposits most of the laser energy into the nozzle and drives the quasistatic axial field.

Simulations are performed in 2D Cartesian geometry ($200 \times 40\ \mu\mathrm{m}^2$, $100\ \mathrm{cells}/\mu\mathrm{m}$, $t_{\mathrm{end}} = 1$ ps), with fully ionized Al ($n_{\mathrm{Al}} = 6 \times 10^{22}\ \mathrm{cm}^{-3}$) for the nozzle and fully ionized hydrogen ($n_{\mathrm{H}} = 5 \times 10^{22}\ \mathrm{cm}^{-3}$) for the rod.

### B. Analytical model of source-field synchronization

The prepulse ($I_{\mathrm{pre}} = 5 \times 10^{21}\ \mathrm{W/cm}^2$, $a_0 \simeq 48$) heats electrons to a ponderomotive temperature [3] $T_h = m_e c^2\left(\sqrt{1 + a_0^2} - 1\right) \simeq 24$ MeV, driving a proton front at $v_{\mathrm{front}} \approx 2\, c_s \simeq 0.3\, c$, where $c_s = \sqrt{T_h/m_p}$ and the factor of 2 follows from the early-time self-similar expansion [35].

Both pulses are assumed to originate from $x = 0$ (laser injection point). The prepulse propagates at $c$ to the H-rod at $x_{\mathrm{source}}$, extracts protons, which then fly a distance $d = x_{\mathrm{skirt}} - x_{\mathrm{source}}$ to the nozzle acceleration zone at $v_{\mathrm{front}}$. The main pulse reaches $x_{\mathrm{skirt}}$ at the speed of light. Equating arrival times gives the synchronization condition:

$$\Delta t_{\mathrm{opt}} = d\left(\frac{1}{v_{\mathrm{front}}} - \frac{1}{c}\right), \qquad (1)$$

i.e., the difference between the proton and photon transit times over the same distance $d$. Tracking the on-axis $E_x$

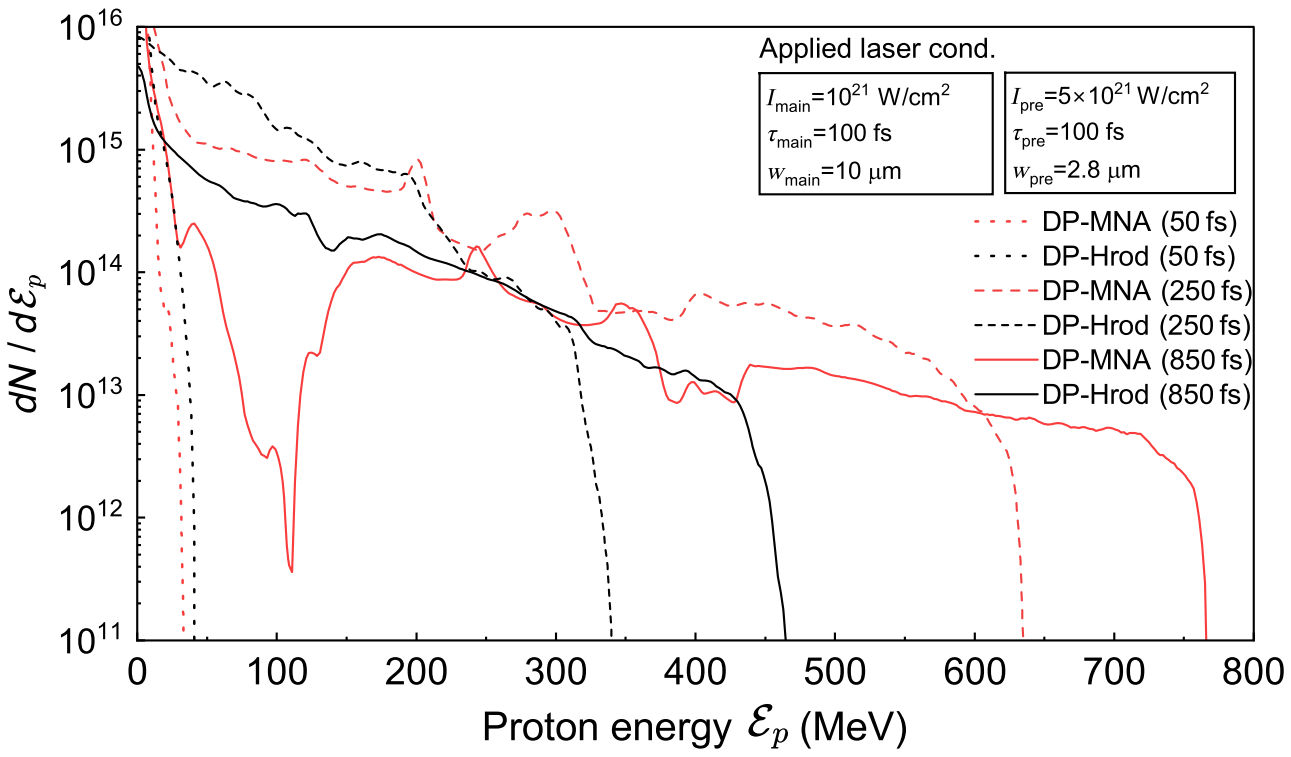

FIG. 2. Proton energy spectra for DP-MNA (red) and DP–H-rod (black) under identical laser conditions at representative times. DP–MNA develops a harder high-energy tail extending toward the sub-GeV regime.

peak in the PIC data reveals a comoving accelerating structure at $x_{\mathrm{skirt}} \simeq 9\,\mu\mathrm{m}$ ($t \gtrsim 200$ fs), propagating at $\sim 0.25\,c \approx v_{\mathrm{front}}$. For a midsurface proton ($d \simeq 2\,\mu\mathrm{m}$), $\Delta t_{\mathrm{opt}} \simeq 16$ fs—within the optimal band 0–40 fs in Fig. 4.

The ellipsoidal H-rod spans from tip ($d_{\mathrm{tip}} = 1.2\,\mu\mathrm{m}$) to equator ($d_{\mathrm{eq}} = 4.0\,\mu\mathrm{m}$), yielding $\Delta t_{\mathrm{tip}} \simeq 9$ fs and $\Delta t_{\mathrm{eq}} \simeq 31$ fs. The synchronization window width is therefore

$$\delta(\Delta t) = r_x \left( \frac{1}{v_{\mathrm{front}}} - \frac{1}{c} \right) \simeq 22 \text{ fs}, \tag{2}$$

where $r_x = 2.8\,\mu\mathrm{m}$ is the H-rod semiaxis. This intrinsic $\sim$22 fs tolerance, supplemented by the finite extent of the acceleration region ($L_{\mathrm{acc}} \sim 10\ \mu\mathrm{m}$), accounts for the $\sim$40 fs bandwidth in Fig. 4. Equation (2) provides a design rule: the window scales linearly with source size and inversely with front velocity.

The cutoff energy scaling $\mathcal{E}_{p,\max} \propto T_e^{3/2} \propto I^{3/4}$ follows from the MNA field model [28]: the nozzle geometry sets the acceleration length to the Debye length $\lambda_{De} \propto T_e^{1/2}$ rather than the density scale height characteristic of TNSA ($\mathcal{E}_p \propto T_e \propto I^{1/2}$), consistent with the simulation fit $\mathcal{E}_{p,\max} \propto I_L^{0.79}$ reported therein and with the sublinear trend in Fig. 5.

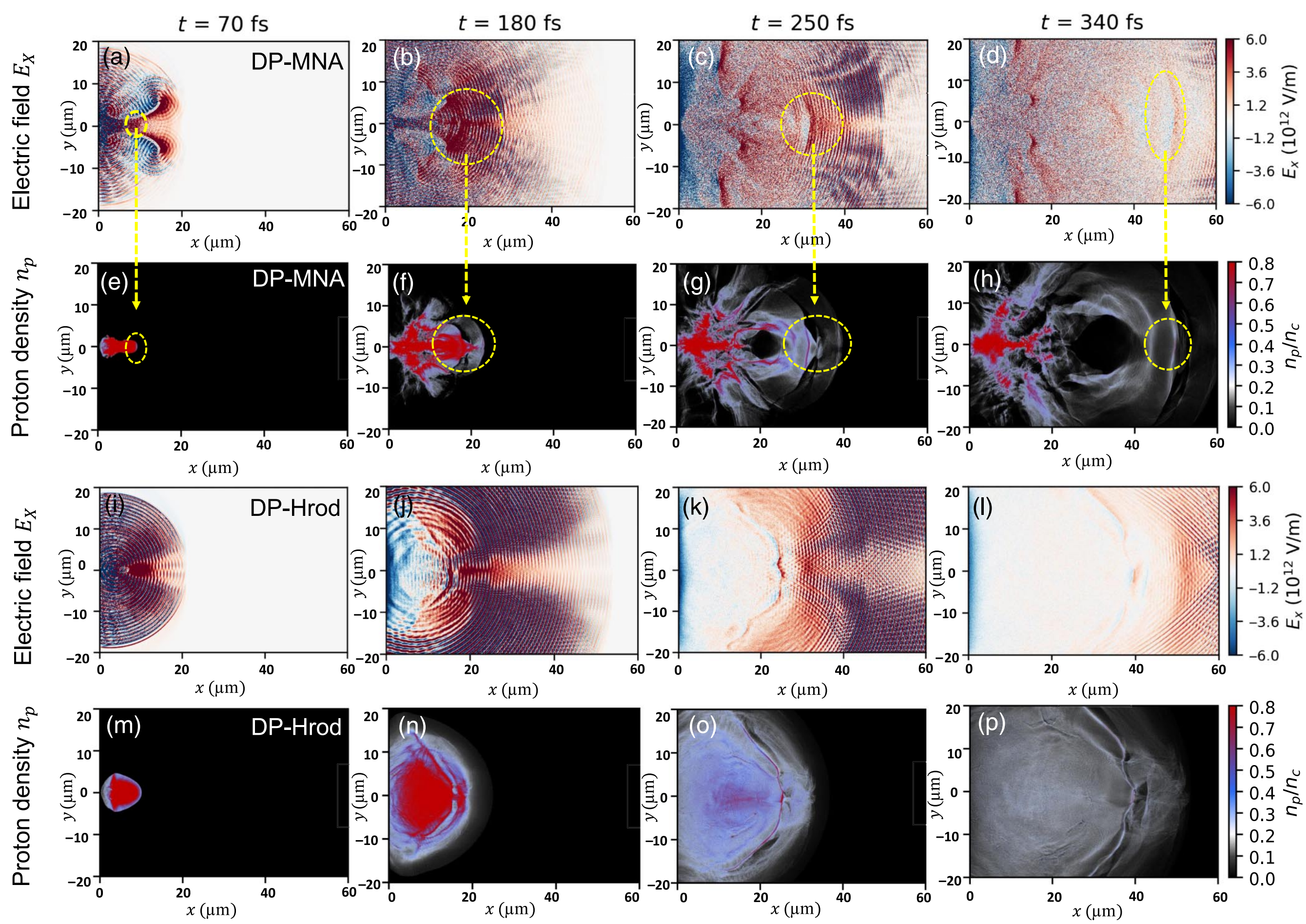

FIG. 3. Evolution of $E_x$ and $n_p$ for DP-MNA and DP–H-rod. (a)–(d) $E_x$ for DP-MNA; (e)–(h) $n_p$ for DP-MNA; (i)–(l) $E_x$ for DP–H-rod; (m)–(p) $n_p$ for DP–H-rod. DP-MNA sustains a comoving axial channel phase-locked to the proton front; DP–H-rod exhibits rapid field decay and debunching.

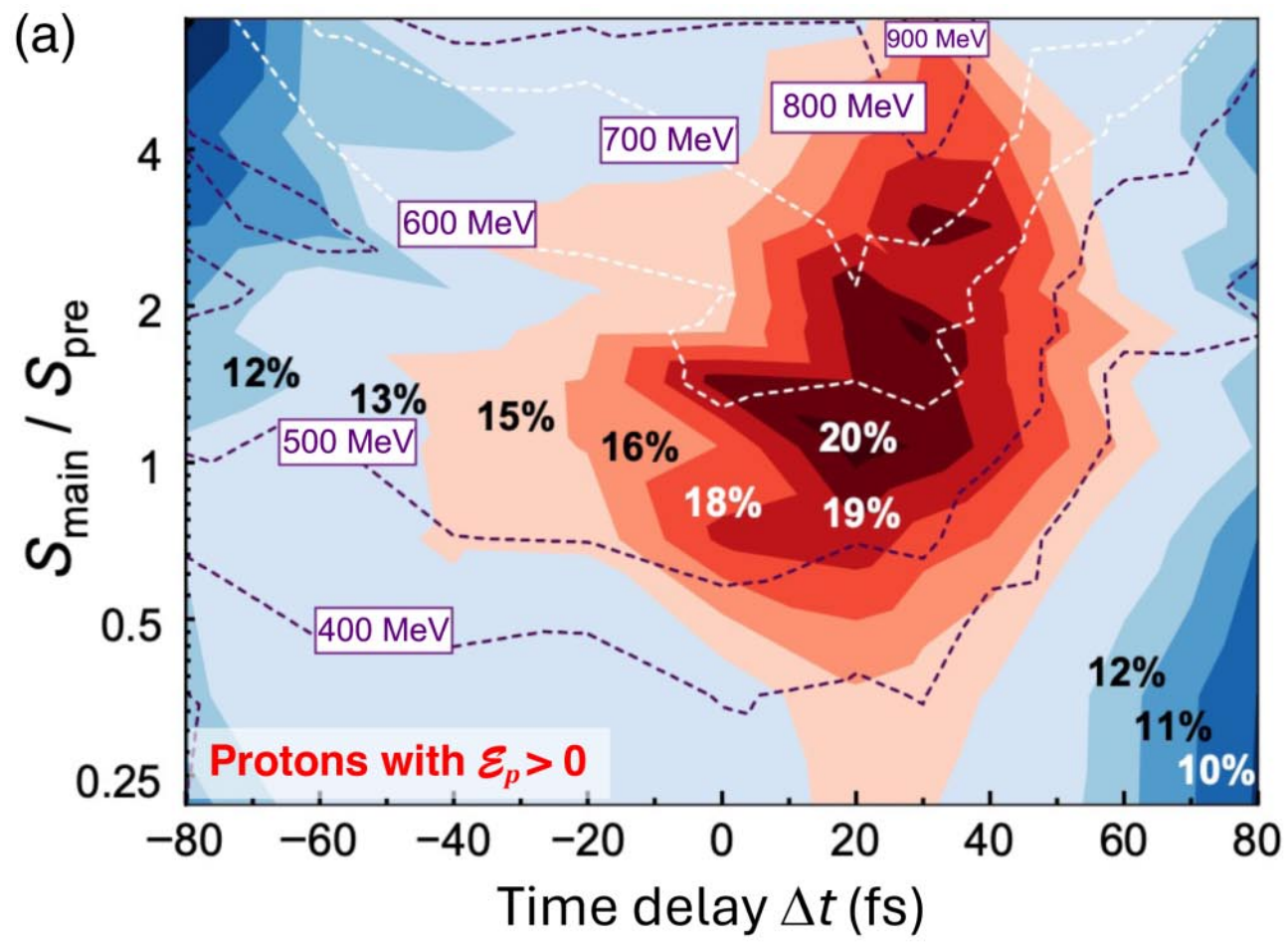

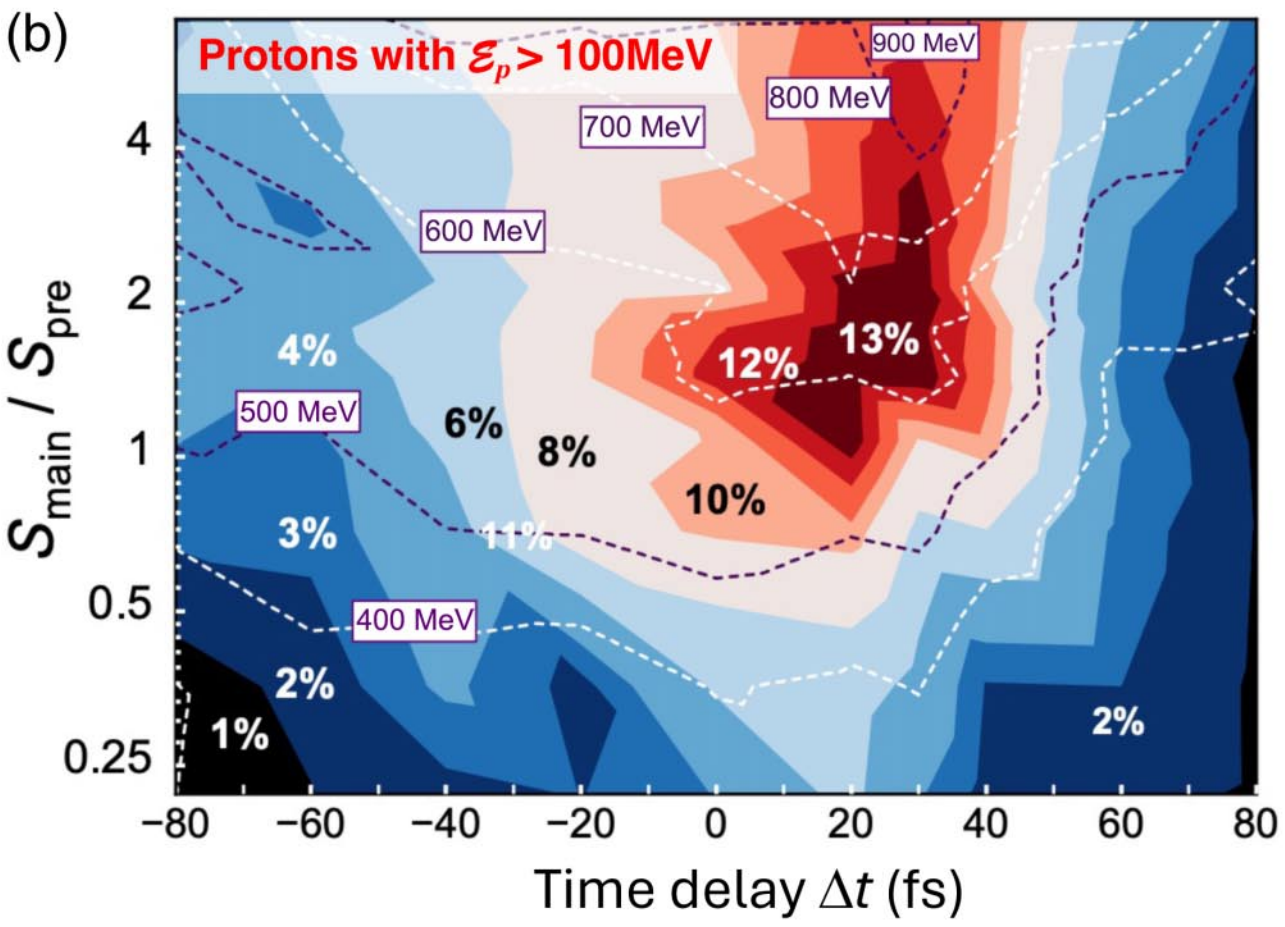

FIG. 4. (a) Total efficiency $\eta$ and (b) tail efficiency $\eta_{>100}$ versus $\Delta t$ and $S_{\mathrm{main}}/S_{\mathrm{pre}}$. Both peak at $\Delta t \simeq 0$–40 fs.

### C. Phase-locked acceleration: 2D results

To isolate the nozzle-cavity effect under dual-pulse drive, we compare DP-MNA with DP–H-rod (without nozzle confinement) under identical laser parameters (Fig. 2). The spectra correspond to $\Delta t \simeq 30$ fs. At an early time ($t = 50$ fs), the proton spectra are nearly indistinguishable. From $t = 250$ fs onward, DP-MNA develops a broad, slowly decaying high-energy tail: for $\mathcal{E}_p \gtrsim 200$ MeV, the spectral density exceeds DP–H-rod by factors of a few and remains substantial beyond 500 MeV, whereas DP–H-rod retains a quasiexponential decay terminating above $\sim$350 MeV. By $t = 850$ fs, the DP-MNA cutoff extends to $\gtrsim$760 MeV while DP–H-rod remains limited to $\sim$460 MeV.

These trends show that the performance gain arises not from the mere presence of a second pulse but from the spatiotemporal synergy between the dual-pulse drive and the source-field-separated geometry. Importantly, this sub-GeV cutoff is achieved at $I_{\mathrm{main}} = 10^{21}$ W/cm$^2$, indicating that geometry-assisted temporal control can approach sub-GeV-class energies without brute-force intensity scaling.

Figure 3 links this spectral divergence to the spatiotemporal evolution of $E_x$ and the proton density. In DP-MNA, a narrow comoving axial channel with a lifetime of several hundred femtoseconds is maintained; the dense proton front remains embedded and gains energy over an extended distance. In DP–H-rod, the axial field collapses within a few hundred femtoseconds and the proton front debunches, terminating the acceleration at lower energies.

### D. Delay window and efficiency-energy trade-off

Guided by the phase locking in Fig. 3, we scan $\Delta t$ and the energy sharing $S_{\mathrm{main}}/S_{\mathrm{pre}}$, where $S \equiv I\tau w$ parametrizes the invested pulse energy in 2D. The prepulse is fixed at $I_{\mathrm{pre}} = 5 \times 10^{21}$ W/cm$^2$, $\tau_{\mathrm{pre}} = 100$ fs.

Figures 4(a) and 4(b) map $\eta$ and $\eta_{>100}$ over ($\Delta t$, $S_{\mathrm{main}}/S_{\mathrm{pre}}$). A distinct high-performance band appears at $\Delta t \simeq 0$–40 fs, confirming synchronization timing as the dominant control parameter. Within this band, $\eta \simeq 20\%$ and $\eta_{>100} \simeq 12\%$–13%, implying $\eta_{>100}/\eta \sim 0.65$: a majority of the converted energy resides in the >100 MeV component rather than a low-energy bulk. Such a high above-threshold fraction is incompatible with quasithermal sheath expansion and indicates preferential energy loading into a phase-locked population.

Figure 5 projects the synchronized cases onto $(\mathcal{E}_{p,\mathrm{max}}, \eta_{>100})$. The data show $\eta_{>100} \gtrsim 10\%$ is maintained as $\mathcal{E}_{p,\mathrm{max}}$ approaches 0.8–1.0 GeV, alleviating the usual trade-off. $\mathcal{E}_{p,\mathrm{max}}$ is advanced primarily by $I_{\mathrm{main}}$ (marker size), while $\eta_{>100}$ peaks at $\tau_{\mathrm{main}} \approx 60$–100 fs (color), consistent with transit-time matching to the cavity-field lifetime.

### E. Numerical convergence

Convergence tests were performed with respect to spatial resolution, particle statistics, and domain size using the baseline delay condition ($\Delta t \simeq 30$ fs). Table I summarizes the results. The cutoff is insensitive to particle count ($\Delta < 3\%$) and domain size ($\Delta < 3\%$). The 50 cells/μm case underestimates the cutoff by ∼29%, while the 200 cells/μm run agrees to within 5%; the baseline resolution of 100 cells/μm is therefore well converged. The conversion efficiencies $\eta$ and $\eta_{>100}$ are equally well converged, with $\eta_{>100}$ varying by less than 2% points across all adequately resolved cases.

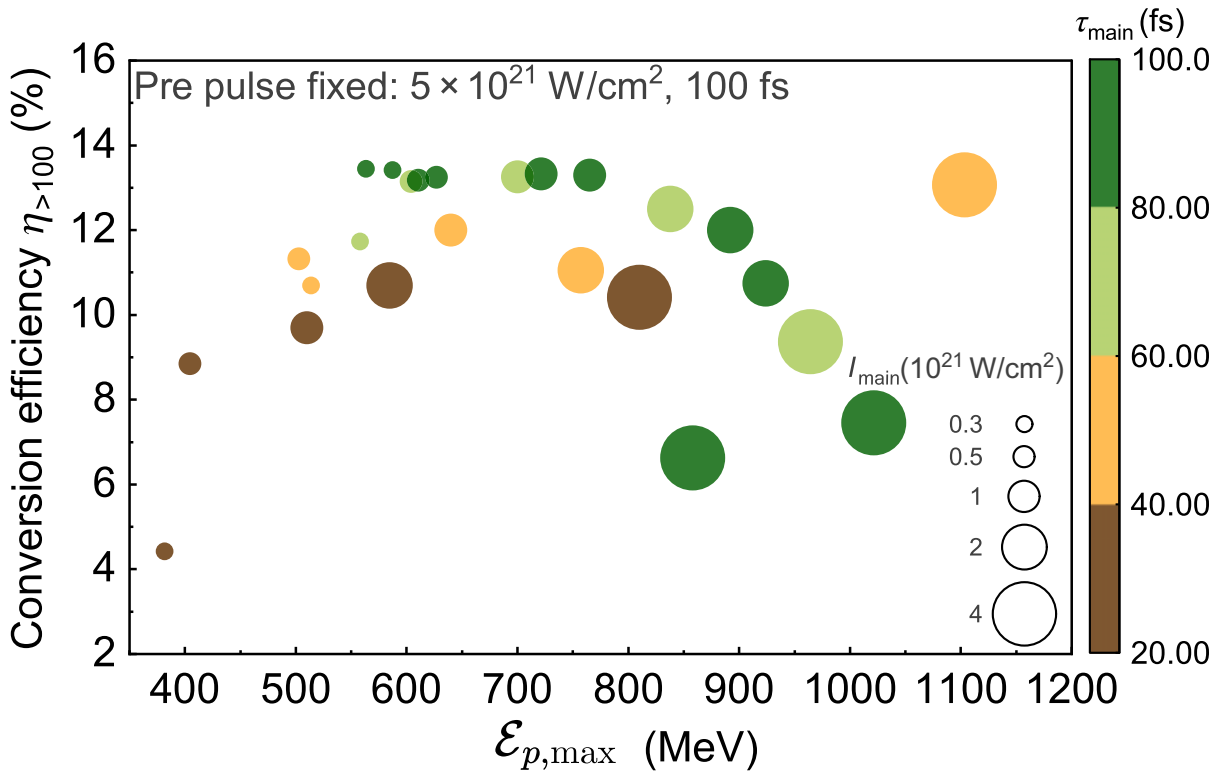

FIG. 5. Synchronized cases on the $(\mathcal{E}_{p,\mathrm{max}}, \eta_{>100})$ plane. Marker size $\propto I_{\mathrm{main}}$; color encodes $\tau_{\mathrm{main}}$.

TABLE I. Convergence of the proton cutoff energy $\mathcal{E}_{p,\max}$. The baseline case is marked in bold.

| Test | Parameters | $\mathcal{E}_{p,\max}$ (MeV) | Δ |
| --- | --- | --- | --- |
| Resolution | 50 cells/μm | 546 | −29% |
| | **100 cells/μm** | **769** | ⋯ |
| | 200 cells/μm | 734 | −5% |
| PPC | **100/100/200** | **769** | ⋯ |
| | 200/200/400 | 784 | +2% |
| Domain | $150 \times 30$ μm$^2$ | 753 | −2% |
| | $200 \times 40$ μm$^2$ | **769** | ⋯ |
| | $250 \times 50$ μm$^2$ | 746 | −3% |

### F. Three-dimensional validation

To assess the robustness of the phase-locking mechanism beyond the 2D approximation, we perform 3D PIC simulations using a slit-nozzle geometry: the 2D cross section is extruded along $z$ over a finite length $L_z$, forming two parallel Al plates that confine the proton channel in $y$ while leaving $z$ open at the plate edges. This geometry is compared with an unconfined 3D H-rod (DP–H-rod) under identical laser conditions. All 3D simulations use 30 cells/μm, 10/10/20 PPC (H/Al/e), and a domain of $100 \times 24 \times 24$ μm$^3$.

*Confined versus unconfined target*—Figure 6 compares the proton spectra for DP-MNA ($L_z = 10$ μm) and DP–H-rod at $t = 50$, 250, and 500 fs. As in 2D, the early-time spectra are similar; at later times, DP-MNA develops a harder tail with a peak cutoff of ∼263 versus 167 MeV for DP–H-rod—a 57% enhancement under identical laser conditions. The absolute energies are reduced compared to 2D (as expected from the additional transverse degree of freedom, which enhances electron escape and weakens the axial field), but the relative advantage of the confined geometry is preserved.

*Dual versus single pulse*—Figure 7 shows $\mathcal{E}_{\max}$ versus time. DP-MNA rises steeply and reaches its peak at ∼870 fs, whereas DP-H-rod saturates gradually at ∼970 fs. Also plotted is the single-pulse case (main pulse only, no prepulse), which reaches only ∼21 MeV—confirming that the prepulse-driven proton injection is essential and that nozzle confinement alone does not produce phase-locked acceleration. Both geometric confinement and temporal synchronization are necessary.

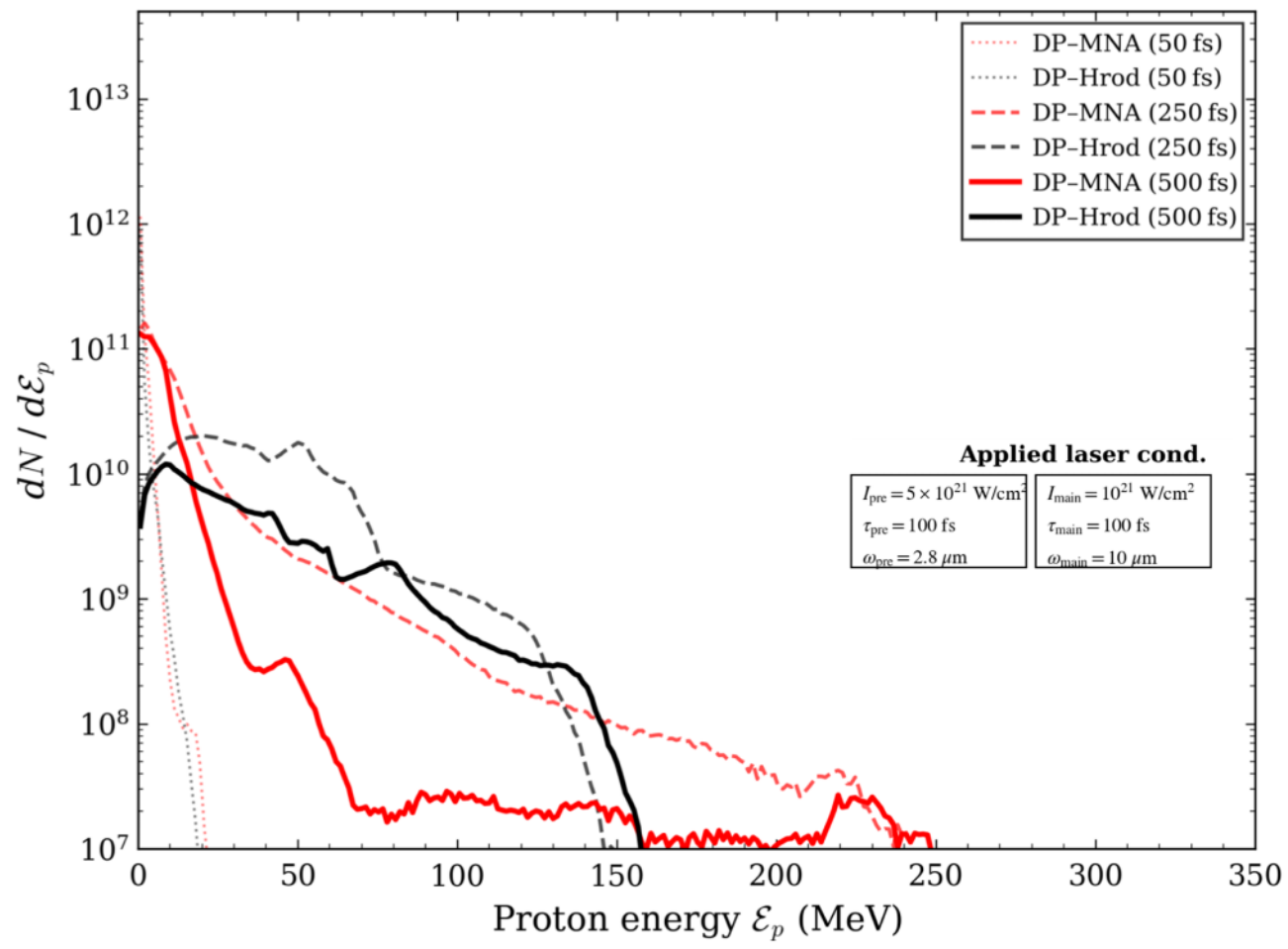

FIG. 6. 3D proton energy spectra for DP-MNA (slit, $L_z = 10$ μm; red) and DP–H-rod (black) at representative times.

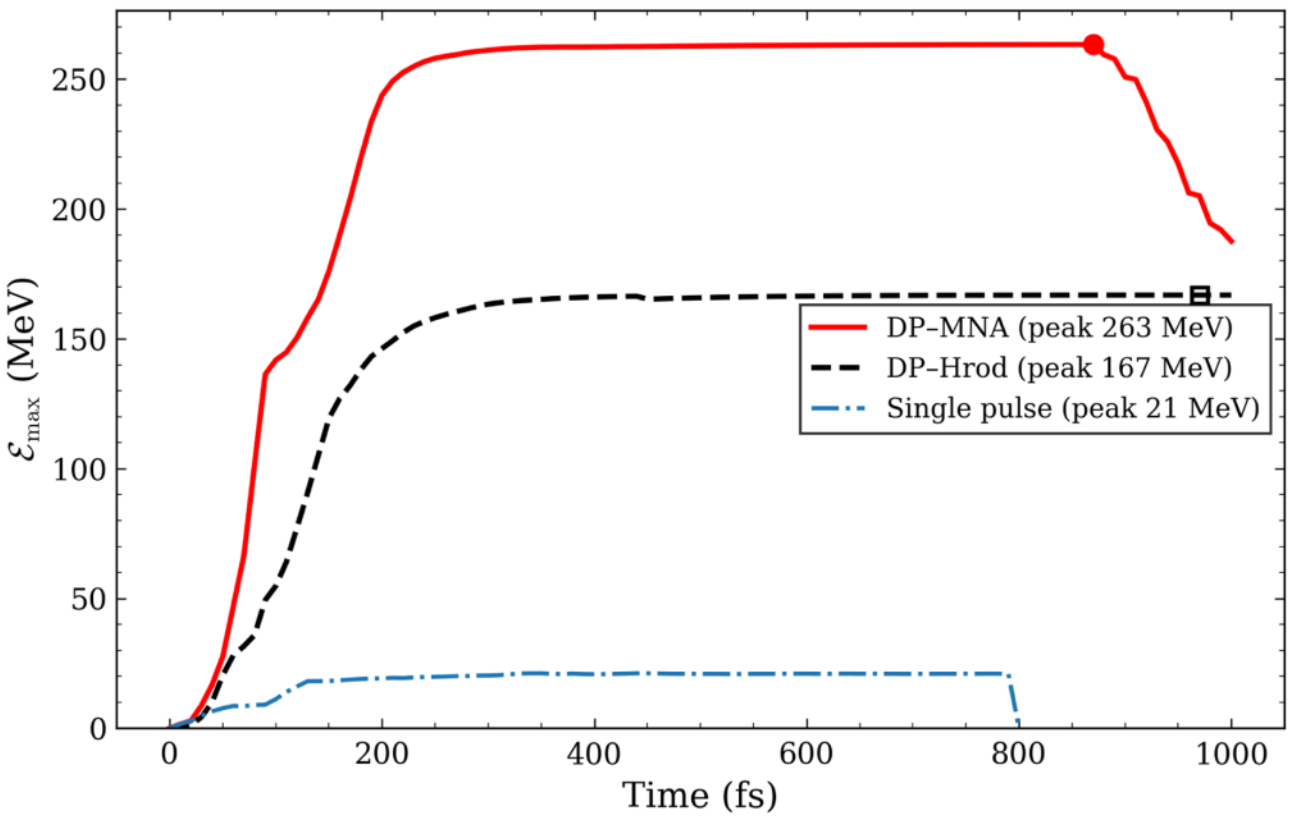

FIG. 7. Maximum proton energy versus time in 3D. DP-MNA (dual pulse + nozzle) reaches 263 MeV; the single-pulse case (nozzle only, no prepulse) yields only 21 MeV, demonstrating that both confinement and temporal synchronization are necessary.

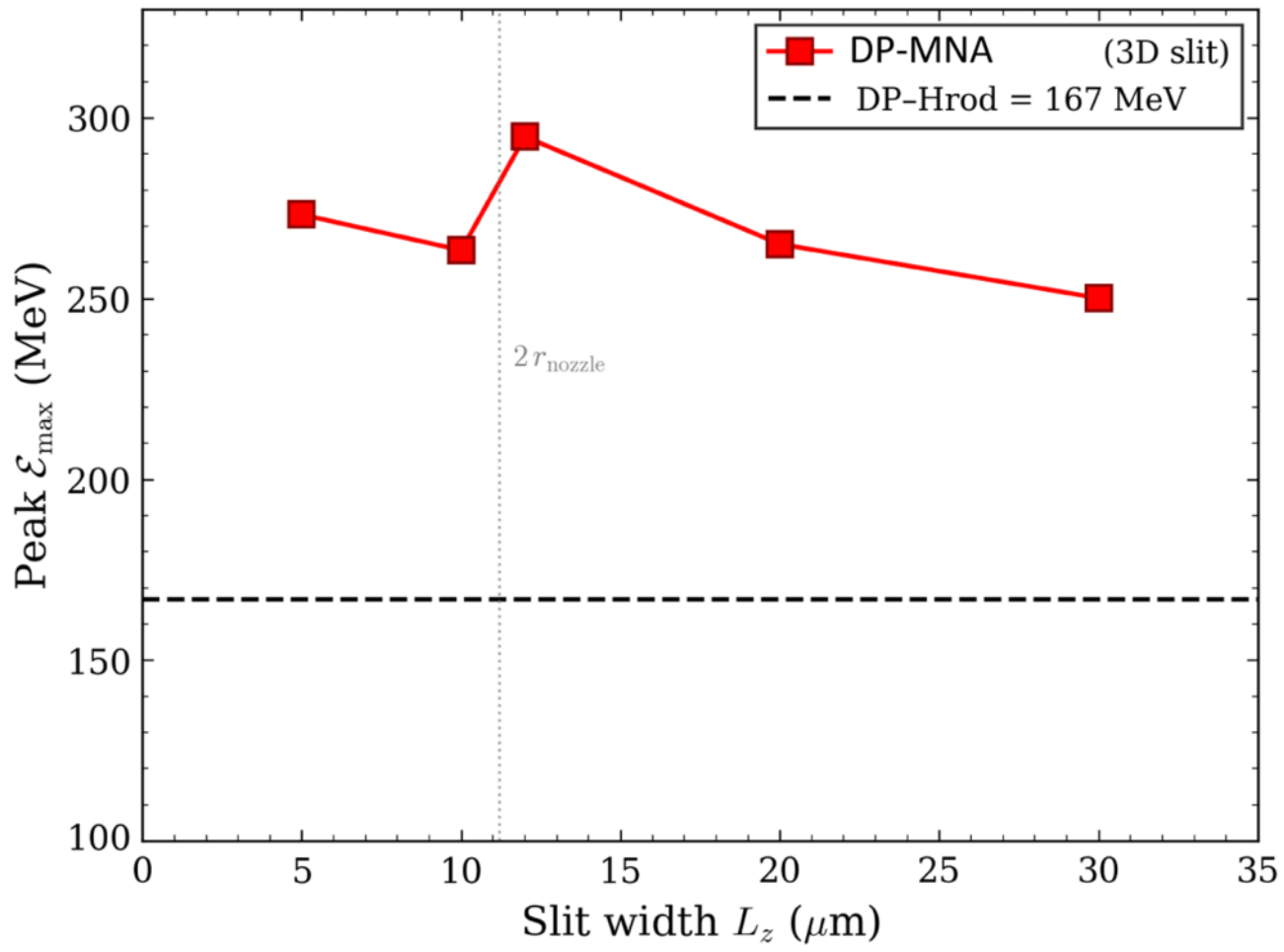

FIG. 8. Peak proton cutoff energy versus slit width $L_z$. All values exceed the DP–H-rod baseline (dashed line, 167 MeV). An optimum near $L_z \simeq 10$–12 μm reflects a balance between laser coupling and transverse confinement.

*Slit-width optimization*—Figure 8 shows the peak $\mathcal{E}_{p,\max}$ as a function of the slit width $L_z$ for $L_z = 5$, 10, 12, 20, and 30 μm. All values exceed the DP–H-rod baseline of 167 MeV. An optimum near $L_z \simeq 10$–12 μm—close to twice the nozzle inner radius ($2\,r_{\mathrm{nozzle}} \simeq 11.2$ μm)—reflects a balance between laser-energy coupling (which favors larger $L_z$) and electron confinement in the $z$ direction (which favors smaller $L_z$). This trade-off is an inherently 3D effect with no counterpart in the 2D model.

*Delay scan in 3D*—A scan over $\Delta t = 0$, 20, 30, and 80 fs at fixed $L_z = 10$ μm yields peak cutoffs of 247, 257, 263, and 216 MeV, respectively. The monotonic decrease with increasing $\Delta t$ is consistent with the analytical synchronization window [Eqs. (1) and (2)], confirming that the delay sensitivity persists in three dimensions.

## III. SUMMARY

In summary, DP-MNA alleviates the efficiency-energy trade-off by enforcing source-field synchronization in a micronozzle cavity. A delay-defined window locks a prepulse-bunched proton front to a comoving quasistatic axial field, preserving bunch integrity and extending the effective acceleration stage. At $10^{21}$ W/cm$^2$-class intensities, we obtain sub-GeV-class cutoffs with $\eta \sim 20\%$, while the $\mathcal{E}_p > 100$ MeV component alone reaches $\eta_{>100} \sim 13\%$—signifying direct energy loading into a compact, application-relevant population.

An analytical model [Eq. (1)] explains the optimal delay as the difference between the proton and photon travel times to the nozzle skirt, with the synchronization window width set by the H-rod size [Eq. (2)]. Convergence tests confirm the reliability of the 2D simulation parameters (Table I).

Three-dimensional PIC simulations using a slit-nozzle geometry verify that the key signatures of phase-locked acceleration—spectral hardening, extended acceleration, and the nozzle-confinement advantage—are preserved in 3D, with the confined target yielding cutoff energies ∼60% higher than an unconfined hydrogen rod. A single-pulse comparison (21 versus 263 MeV) demonstrates that the dual-pulse temporal control is essential: nozzle confinement alone does not activate phase-locked acceleration. A scan over the slit width reveals an optimal $L_z$ near the nozzle diameter, an inherently 3D design parameter. These results establish phase-locked acceleration as a robust design principle for compact, high-yield proton drivers and identify femtosecond timing control and three-dimensional geometry optimization as practical levers for future experimental realization.

## ACKNOWLEDGMENTS

This work was supported by (1) Kansai Electric Power Company, Incorporated (KEPCO) and (2) D3 Center, Osaka University (computing resources via supercomputer SQUID).

## DATA AVAILABILITY

The data that support the findings of this article are not publicly available upon publication because it is not technically feasible and/or the cost of preparing, depositing, and hosting the data would be prohibitive within the terms of this research project. The data are available from the authors upon reasonable request.